\documentclass[aps,prd,10pt,twocolumn,showpacs,preprintnumbers,superscriptaddress, nofootinbib]{revtex4-1}
\usepackage{amsmath,amssymb,amsfonts,amsthm}
\usepackage{amsbsy} 
\usepackage{epsfig}
\usepackage{hyperref}
\usepackage{bm}

\usepackage{color}

\renewcommand{\eqref}[1]{(\ref{#1})}
\newcommand{\ba}{\begin{eqnarray}}
\newcommand{\ea}{\end{eqnarray}}

\def\ie{i.e., }
\def\eg{e.g., }

\def\Ds{D_{\textsc s}}
\def\ds{d_{\textsc s}}

\def\dh{d_{\textsc h}}
\def\dw{d_{\textsc w}}

\def\e{e}

\def\d{d}
\def\kmax{k_{\rm{max}}}

\renewcommand{\digamma}{\psi}

\newcommand{\std}{D}				% space-time dimension
\newcommand{\sd}{d}				% spatial dimension

\newcommand{\cm}{\mathcal{C}}
\newcommand{\cminf}{{\mathcal{C}_\infty}}
\newcommand{\p}{p}					% dimension of cells, forms etc
\newcommand{\size}{N}				% "lattice" size 	
\newcommand{\qs}{\psi}				% generic quantum state			
\newcommand{\qsc}{a}	
\newcommand{\jmin}{{j_{\text{min}}}}
\newcommand{\jmax}{{j_{\text{max}}}}
\newcommand{\rjc}{|\{j_c\},\cm\rangle}
\newcommand{\ljc}{\langle \{j_c\},\cm |}
\newcommand{\rj}{|j,\cm\rangle}
\newcommand{\lj}{\langle j,\cm |}
\newcommand{\rjb}{|j,\cm_\size \rangle}
\newcommand{\ljb}{\langle j,\cm_\size |}
\newcommand{\rsup}{| V_0,\jmin,\jmax \rangle}
\newcommand{\ssup}{{V_0,\jmin,\jmax}}

\newcommand{\ra}{\rightarrow}

\newcommand{\os}{\overset}
\newcommand{\In}{\subset}

\renewcommand\[{\begin{equation}}
\renewcommand\]{\end{equation}}

\newcommand{\R}{\mathbb R}
\newcommand{\Q}{\mathbb Q}
\newcommand{\Z}{\mathbb Z}
\newcommand{\N}{\mathbb N}

\newcommand{\Tr}{\mathrm{Tr}}

\newcommand{\sumint}{\sum}
\def\com{\color{magenta}}
\def\cob{\color{blue}}
\newcommand{\au}[2]{#1.\ #2}
\newcommand{\books}[4]{\emph{#1} (#2, #3, #4)}
\newcommand{\oarX}[1]{\href{http://arxiv.org/abs/#1}{{\ttfamily\com arXiv:#1}}}
\newcommand{\arX}[1]{\href{http://arxiv.org/abs/#1}{{\ttfamily\com arXiv:#1}}}
\newcommand{\doin}[6]{\href{http://dx.doi.org/#1}{{\cob #2 #3 {\bf #4}, #5 (#6)}}}
\newcommand{\doinn}[5]{\href{http://dx.doi.org/#1}{{\cob #2 {\bf #3}, #4 (#5)}}}
\newcommand{\doij}[5]{\href{http://dx.doi.org/#1}{{\cob #2 #3 (#5) #4}}}

\newcommand{\procsin}[5]{in \emph{#1}, edited by #2 (#3, #4, #5)} % for Singapore
\newcommand{\tia}[1]{}

\newcommand{\COTa}{Calcagni:2013ku}
\newcommand{\COTb}{Calcagni:2014ep}

\newcommand{\lqgT}{Thiemann:2007wt}
\newcommand{\lqgR}{Rovelli:2004wb,Rovelli:2011tk}%,Rovelli:2014vg}
\newcommand{\sfOP}{Perez:2003wk,Perez:2013uz}
\newcommand{\gftFO}{Freidel:2005jy,Oriti:2012wt}%,Oriti:2006ts}
\newcommand{\cdtA}{Ambjorn:2012vc}%,Ambjorn:2000hp}

\newcommand{\asR}{Niedermaier:2006up,Reuter:2012jx}
\newcommand{\hlH}{Horava:2009ho}

\newcommand{\cdtfractal}{Ambjorn:2005fj,Ambjorn:2005fh}
\newcommand{\asfractal}{Lauscher:2005kn}
\newcommand{\hlfractal}{Horava:2009ho}
\newcommand{\ncfractal}{Benedetti:2009fo,Alesci:2012jl,Arzano:2014ke}
\newcommand{\snfractal}{Modesto:2009bc}

\begin{document}

\title{Dimensional flow in discrete quantum geometries}

\author{Gianluca Calcagni}
\email{calcagni@iem.cfmac.csic.es}
\affiliation{Instituto de Estructura de la Materia, CSIC, Serrano 121, 28006 Madrid, Spain}
\author{Daniele Oriti}
\email{doriti@aei.mpg.de}
\author{Johannes Th\"urigen}
\email{johannes.thuerigen@aei.mpg.de}
\affiliation{Max Planck Institute for Gravitational Physics (Albert Einstein Institute)\\
Am M\"uhlenberg 1, D-14476 Potsdam, Germany}

\date{\small December 29, 2014}

\begin{abstract}
In various theories of quantum gravity, one observes a change in the spectral dimension from the topological spatial dimension $d$ at large length scales to some smaller value at small, Planckian scales. While the origin of such a flow is well understood in continuum approaches, in theories built on discrete structures a firm control of the underlying mechanism is still missing. We shed some light on the issue by presenting a particular class of quantum geometries with a flow in the spectral dimension, given by superpositions of states defined on regular complexes. For particular superposition coefficients parametrized by a real number $0<\alpha<d$, we find that the spatial spectral dimension reduces to $\ds \simeq \alpha$ at small scales. The spatial Hausdorff dimension of such class of states varies between 1 and $d$, while the walk dimension takes the usual value $\dw=2$. Therefore, these quantum geometries may be considered as fractal only when $\alpha=1$, where the ``magic number'' $\Ds\simeq 2$ for the spectral dimension of space\emph{time}, appearing so often in quantum gravity, is reproduced as well. These results apply, in particular, to special superpositions of spin-network states in loop quantum gravity, and they provide more solid indications of dimensional flow in this approach.
\end{abstract}

\preprint{\doin{10.1103/PhysRevD.91.084047}{PHYSICAL REVIEW}{D}{91}{084047}{2015} \hspace{9cm} \arX{1412.8390}}
\preprint{AEI-2014-028}

\maketitle

% Introduction ==================================================================

\section{Introduction} 

% Background
The identification of good geometric observables is a thorny issue in (quantum) gravitational physics, and it is of particular importance in nonperturbative, background-independent approaches to quantum gravity, especially where the fundamental degrees of freedom characterizing quantum states and histories of the system are nongeometric in the standard sense and characterized by intrinsic discreteness. Examples are loop quantum gravity (LQG) \cite{\lqgT,\lqgR}, spin-foam models \cite{\sfOP} and group field theory (GFT) \cite{\gftFO}, strictly related to LQG \cite{Oriti:2014wf,Oriti:2014td}. Here, the major challenge is to find a relation to the continuum spacetime geometries of classical general relativity, \ie to show that the latter emerge from the fundamental discrete quantum structures of the theory in some approximation. This emergence has to be expressed in terms of suitable geometry observables, both classical and quantum, that should indicate that the desired features of smooth spacetimes are recovered. This is, in fact, a precondition for extracting physics from such quantum-gravity formalisms. %Both classical and quantum observables are an essential ingredient to understand this relation.

Effective-dimension observables provide important information about the geometric properties of quantum states of space and spacetime histories in quantum gravity.
In particular, the spectral dimension $\ds$, which depends on the spectral properties of a geometry through its definition as the scaling of the heat-kernel trace, has attracted special attention due to the observation of a dimensional flow (i.e., the change of spacetime dimensionality across a range of scales \cite{tHo93,Car09,fra1}) in various approaches, such as causal dynamical triangulations (CDT) \cite{\cdtA}, the functional renormalization-group approach of asymptotic safety \cite{\asR} and Ho\v{r}ava--Lifshitz gravity \cite{\hlH} among others. 

In all these approaches, the spectral dimension of spacetime exhibits a scale dependence itself, flowing from the topological dimension $\std$ in the infrared (IR) to $\Ds\simeq 2$ in the ultraviolet (UV) \cite{Ambjorn:2005fj,Ambjorn:2005fh,Lauscher:2005kn,Benedetti:2009bi,Horava:2009ho,CES} (although new CDT calculations \cite{Coumbe:2014wq} rather hint at $\Ds\simeq 3/2$). While modified dispersion relations provide an obvious reason for this behaviour in smooth geometries \cite{\asfractal,\hlfractal,\ncfractal,\snfractal}, dimensional flow remains to be better understood in the case of discrete calculations as in the CDT approach \cite{\cdtfractal,Benedetti:2009bi, Coumbe:2014wq}. Causal dynamical triangulations, in fact, aim at a definition of the continuum path integral for quantum gravity via a regularization of the same in terms of a superposition of simplicial complexes (thus a form of discrete geometries) weighted by the Regge action. While it is more difficult to identify the underlying reason for the dimensional flow in this context, the same is obtained in a very direct manner from the evaluation of the heat trace as a quantum geometric observable inside the CDT partition function. %as in CDT \cjt{technically discrete, but continuum geometries}  {\bf [this claim clashes with the one I inserted in the introduction. Again, CDT is not a theory of discrete geometries but a regularization of smooth ones]} 

Here we take a very similar direct approach, but in a context that it is closer to the formalism of loop quantum gravity. In LQG, quantum states are defined as superpositions of spin networks, which are graphs labeled by algebraic data from the representation theory of $SU(2)$. There is thus an interplay between two types of data and their corresponding discreteness: a combinatorial discreteness due to the graph substratum for the quantum states, and an algebraic discreteness due to the fact that the labels are half-integers corresponding to $SU(2)$ irreducible representations. Quantum effects in the evaluation of observables are thus to be expected, in general, from both these sources and it is an important limitation to focus only on one of them, as in preliminary studies of dimensional flow in LQG \cite{\snfractal}. 

In a previous work \cite{\COTb}, we have already tackled the issue of computing the spectral dimension on LQG states based on a given graph, dealing both with coherent states and with their superpositions. There, we showed that the underlying discrete structure plays a dominant role. Here, we intend to explore the role of combinatorial discreteness and of superpositions of combinatorial structures in greater detail. 

% New in this work
In this paper, we present a special class of superpositions of discrete quantum states characterized by a real-valued parameter $\alpha$. This parameter will control the scale-dependent values taken by the spectral dimension, and therefore the dimensional flow. %that $\ds$ takes at different lengths or energies. 
These superpositions are over states based on regular complexes corresponding to hypercubic lattices to which a single quantum label is assigned, uniformly to all cells of a certain dimension. Such states occur indeed in the kinematical Hilbert space of the quantum gravity formalisms we just mentioned: LQG, spin-foam models and GFT. 
Because of the uniform labeling, these superpositions are also similar to the discrete geometries in CDT, although we understand the former not as regularization tools for physically smooth geometries but as fundamentally discrete structures with their own physical interpretation. Contrary to the CDT setting, we interpret the combinatorial structures we superpose as defining quantum gravity states, not histories, and the coefficients in the superpositions to have no immediate dynamical content. However, we point out that this interpretation enters only minimally in the actual calculations and it could be generalized. 

Perhaps surprisingly, superpositions of quantum states supported on different complexes have not been considered much in the LQG literature so far. Instead, most analyses have involved only states based on one and the same complex. A first simple example of states based on superpositions of combinatorial structures are the condensate states with a homogeneous cosmology interpretation introduced recently in the GFT context \cite{GOS1,GOS2,Gie13,Cal14,GO}.

Using a known analytic expression for the spectral dimension of single members in the superposition \cite{\COTb}, we compute numerically superpositions over up to $10^6$ discrete geometries. On these grounds, we find strong evidence for a dimensional flow, characterized by the parameter $\alpha$.

Similarly, we find analytic solutions for the walk dimension and Hausdorff dimension of lattice geometries and perform again numerical calculations of superpositions. 
For these observables, however, while we recover the topological dimension at large scales, we do not find any special properties for superpositions as compared to states defined on fixed complexes.

\section{A general class of superposition states}

% Description ==================================================================

Let us now explain in detail the construction of superposition states of interest, and the calculation of their spectral, walk and Hausdorff dimension.

%\subsubsection*{A general class of superposition states}

%\emph{Discrete quantum geometries} are quantum states whose expectation values of geometric observables constitute discrete geometries.
%A (classical) discrete geometry is a polyhedral complex attached with $\p$-volumes to all its cells of dimension $\p$.
%Usually not all $\p$-volumes are independent.
%%Thereby some $\p$-volumes might be functions of some other $\p'$-volumes. 
%On a simplicial complex, for example, a set of edge lengths ($\p=1$) determines all higher dimensional volumes %
%\footnote{For well-defined, i.e. strictly positive volumes, the edge length have to fulfill generalized triangulation inequalities.}.
%
%
%In a quantum theory the $\p$-volumes become quantum operators acting on states supported on the complex. 
%Thus, their expectation values of these $\p$-volume operators provide a discrete geometry.

%Note that among the class of global observables, defined independently of a specific complex, there are also purely combinatorial operators. For example the number operator of vertices, edges or higher cells of the complex.

Most generally speaking, a discrete quantum state of geometry $|\{j_c\},\cm\rangle$ is given by an assignment of quantum numbers $j_c$ to a certain subset of cells $c\in\cm$ of a (combinatorial) complex $\cm$, 
%and based on combinatorial $\sd$-manifolds $\cm$ (that is polyhedral $\sd$-complexes that are furthermore dimensionally homogenous, strongly connected and non-branching \cite{\ORT}), 
diagonalizing volume operators of these cells
\[\label{spectra}
\widehat {V_{c'}^{(\p)} } \rjc \propto l^\p(j_{c'}) \rjc \,,
\]
where we have adopted natural units.
%We thus interpret these labels as intrinsic metric variables, thus diagonalizing all (commuting) volume operators for cells of the complex dual to the graph on which the quantum states are defined.
An example of such states is spin-network states in LQG, based on the 1-skeleton of the dual complex $\cm^{\star}$, with the $j$'s identifying irreducible representations of $SU(2)$. % (though still we label the states by the primal complex $\cm$ here).
%\subsubsection*{$2+1$-dimensional Euclidean lqg}
In three spacetime dimensions, the spatial ($\sd=2$) states in the spin network basis diagonalize the length operators $\widehat{L}_e$ associated with all edges $e\in\cm$. 
Thus, they are labeled by spins $j_e$ on the corresponding dual edges $e^\star \in\cm^\star$.
The form of the $\widehat{L}_e$ spectra is 
$
l(j_e) = \sqrt{ j_e(j_e+1)+C}
$, 
{with a free parameter $C\in\R$ due to a quantization ambiguity} for the Euclidean theory (as well as for timelike edges in the Lorentzian theory, a continuous positive variable being instead assigned to spacelike edges) \cite{Freidel:2003kx,Achour:2014gr}.
%\[
%\lj \widehat{L_e^2} \rj \propto j(j+1)+c.
%\]
%Thus, summing over these eigen states on a fixed complex $\cm_k=\cm$ amounts to a superposition with $\alpha=1$ asymptotically, and in the case $c=1/4$ exactly (the effect of spins being half-integers can be absorbed into an effective $l_0$, while $k=2j=1,2,...$).

In four spacetime dimensions ($\sd=3$), spin-network states have the same spectrum for area operators $\widehat A_f$ on faces $f\in\cm$ such that \cite{DePietri:1996en,Ashtekar:1997bn}
\[\label{areaspectrum}
l(j_f) = [j_f(j_f+1)+C]^{1/4}\,.
\]
%\[
%\lj \widehat{A_f^2} \rj \propto j(j+1)+c.
%\]
%In both cases most of other geometric quantities can be expressed as functionals of these operators diagonalizing the spin network states \cite{\COTa}.
%Therefore their scaling is captured by $\alpha$ as well.

%As another example, one can consider semiclassical states peaking on local volumes $V^{(p)}$ together with some conjugate variables of extrinsic geometry (which are not relevant here). \cjt{[ still corrupted sentence?]}  These are then labeled directly by their expectation values $j=\langle V^{(p)}\rangle$.

Generic quantum-geometry states are superpositions of the discrete quantum geometries $|\{j_c\},\cm\rangle$, which indeed form a complete spin-network basis of states of the Hilbert space in LQG. 
In particular, this Hilbert space can be cast in the form of a direct sum of Hilbert spaces $\mathcal H = \bigoplus_\cm \mathcal H_\cm$.

% Restriction 1
In the following, we will restrict to superpositions with nonzero coefficients only for states $\rj$ %of quantum geometry
labeled by a single quantum number $j_c = j$ for all cells. %$c\in\cm$. %(for some $0\le\p\le\sd$)
Thus, one can consider the individual states $\rj$ as corresponding to equilateral lattices.
Given this class of quantum states, we then consider generic superpositions of the form %There are two kinds of such superpositions over a class of complexes $\cm$ and over the range of quantum labels $j$ (which may be discrete or continuous, but are assumed to be strictly positive)
\[\label{sups}
|\qs\rangle = \sumint_{j,\cm} \qsc_{j,\cm} \rj \,.
\]
We also impose a constraint on the overall volume $V_0$ computed from such superposition states:
%Either a sum over states with fixed underlying complex $\cm'$, 
%{\bf [we should explain the naive symbol $\delta$ in both expressions or announce that we are going to explain it below, as the pedantic reader might not like it]}
%\[\label{eq2}
%|\qs,\cm'\rangle %= \sumint_{j,\cm} \qsc(j) \delta(\cm,\cm') \rj 
%= \sumint_j \qsc_j |j,\cm' \rangle,
%\]
%or with a fixed global observable such as, for example, an overall volume $V_0$,
\[\label{fixedvolume}
|\qs,V_0\rangle = \sumint_{j,\cm} \qsc_{j,\cm} \,\delta(\lj \widehat{V} \rj, V_0) \,\rj \,,
\]
where the delta is a Kronecker delta. 
%Since spectral observables of discrete geometries are very sensitive to the underlying combinatorial structure and we know that the spectral dimension is meaningful only for sufficiently regular complexes \cite{\COTb},
We will further restrict the sum to certain regular complexes, i.e., hypercubic lattices $\cm _\size$ based on the canonical vertex set $\cm_\size^{[0]} :=  (\Z_\size)^\sd$ of $\sd$-tuples of integers modulo $\size$.  % {\bf [explain $N$ in both symbols. Also, it is not clear how the next equation defines the delta]}. 
In this case, the fixed volume condition is explicitly
\[
V_0  = \ljb \widehat V \rjb \propto \size^\sd\, l^\sd(j)\,,
\]
which fixes the lattice size $\size = \size(j)$ for a given $j$ (at least approximately).\footnote{Ratios $l(j_1)/l(j_2)$ for pairs of quantum numbers $j_1$, $j_2$ can be nonrational, so that one should take the integer value (floor function) of $N(j)$. Physically, it is certainly enough to apply Eq.\,\eqref{fixedvolume} in such an approximative way.  Note also that our results are independent of the spacing of quantum numbers in the superposition. Thus, one could as well define the states as sums over only those $j$'s for which $N(j)\in\N$ strictly.}
In general, there are three scales involved in the superposition states 
\[\label{states}
\rsup := \sumint_{j=\jmin}^{\jmax} \qsc_j |j,\cm_{\size(j)} \rangle\,,
\]
when summing over a finite range from  $\jmin$ to $\jmax$:
a minimal length scale $l(\jmin)$, an intermediate scale $l(\jmax)$ and the overall volume size $V_0^{1/\sd} \propto N(\jmin)l(\jmin) = N(\jmax)l(\jmax)$.
Note that a finite volume $V_0$ bounds also possible cutoffs $\jmax$ (since $N$ is a positive integer).

%\cjt{following 2 paragraphs better in discussion?}

One can also consider the limit of noncompact geometries $N(\jmin) \ra \infty$, where all complexes in the superposition of fixed-volume states \eqref{states} converge to the infinite lattice $\cminf$.
%For the choice of hypercubic lattices and for non-compact geometries, the two cases of 
Thus, they are technically the same as superpositions on the fixed complex $\cminf$, although the physical interpretation is different. %and (\ref{fixedvolume}) are technically the same.
%In the first case, the sum is over all geometries on the fixed lattice $\cminf = \lim_{\size\ra\infty}\Z^\sd_\size = \Z^\sd$, 
%while in the second case each complex in the sum over refined lattices for a fixed but infinite volume $V_0$ (understood in the limit $\size\ra\infty$) is, again, the lattice $\cminf$.
Due to the combinatorial simplicity, results of infinite-size calculations can be directly applied to the finite-volume case. 
%Due to the technical equivalence the results apply also to the fixed complex superposition on the infinite lattice $\cminf$.

%Before discussing the calculations of observables, let us briefly expose the reasons for choosing infinite hypercubic lattices.
%First, already for classical discrete geometries the dimension observables have a clear interpretation (as, indeed, dimensions!) only in case of sufficiently regular, lattice-type complexes. The precise structure (whether, \eg, hypercubic or simplicial) does not play a dominant role, on the other hand \cite{\COTb}.
%Second, in an infinite complex there are no boundary or topology effects, thus simplifying the analysis of quantum effects \cite{\COTb}. 
%Third, for infinite lattices analytic solutions to the dimension observables are known for classical geometries, and thus for single states $\rj$ under the assumption \eqref{assumption}. Only in terms of these analytic solutions for single states is it possible to calculate numerically expectation values for very large quantum superpositions, which is exactly the aim in this work.
Having defined our superposition states, we can move on to the evaluation of the geometric observables of interest, \ie dimension estimators.

% Calculation \ds ===============================================================

\section{Evaluation of dimension observables of superposition states}

\subsection{Spectral dimension}

Let the heat kernel $K(x,x';\tau)$ be the solution of the diffusion equation $(\partial_\tau-\Delta_x)K=0$ on a space $X$, with initial condition $K(x,x';0)=\delta(x-x')$, where $\Delta$ is the Laplace operator on $X$. It is a function of the geometry of $X$ via its assigned metric. In the resolution interpretation of \cite{CM1,CMN}, the parameter $\sqrt{\tau}$ and its inverse represent, respectively, the length scale and the resolution at which a geometry is inspected by a pointwise probe deployed at a spatial point $x'$. The trace of the heat kernel over all points is denoted as $P(\tau)=\Tr_X K(x,x';\tau)$ and called ``return probability'' from the traditional but somewhat problematic interpretation in terms of a diffusing process (see \cite{CES,CMN} for a discussion and resolutions of such problems).

While ordinary diffusion takes place on continuous manifolds, the whole setup, and in particular the definition of the Laplace operator, can be generalized to discrete spaces, like (combinatorial) complexes. This was indeed the subject of \cite{\COTa,Thuerigen:2013vc}. 
%The points $x$ can be defined in a discrete context by vertices of a complex $\cm$ or its dual $\cm^\star$. 
%The key geometric quantity for the evaluation of spectral and walk dimensions is, as anticipated, the Laplacian. 
The Laplacian on $\cm$, as a differential operator acting on a field $\phi_a$ on the $\sd$-cells $c_a\in\cm$ (equivalently, on dual vertices), is then \cite{\COTa}
\ba\label{discrete-lap}
-(\Delta_\cm \phi)_a &=& \sum_{b\sim a} (\Delta_\cm)_{ab} (\phi_a - \phi_b)\nonumber\\
&=& \frac1{V^{(\sd)}_a} \sum_{b\sim a} \frac{V^{(\sd-1)}_{ab}} {l^\star_{ab}} (\phi_a - \phi_b)\,,
\ea
where the sum runs over $\sd$-cells $c_b$ adjacent to $c_a$, $V^{(\sd)}_a$ is the volume of the cell $c_a$, $V^{(\sd-1)}_{ab}$ is the volume of the common bounding $(\sd-1)$-cell and $l^\star_{ab}$ is the length of its dual edge. 
Accordingly, the heat trace on $\cm$ is given by a trace $\Tr_\cm$ over maps of that field space.

Both the return probability $P(\tau)$ and the Laplacian $\Delta$ can be turned into operators $\widehat{P(\tau)}$ and $\widehat \Delta$ acting on quantum states of geometry. %\footnote{Note that local observables such as the Laplacian are only defined on states on a given complex (which is further discussed below).} 
Quantizing the metric-dependent coefficients $(\Delta_\cm)_{ab}$ % \ra \widehat{\Delta^\cm_{ab}}$
 which enter in the definition of the discrete Laplacian \eqref{discrete-lap}
  results in an operator $\widehat{\Delta}_\cm$ acting on the Hilbert space $\mathcal H_\cm$ of states on a given complex $\cm$ 
which returns states together with discrete Laplacians.%, \ie maps on the field space of $\cm$.%
\footnote{
%On the other hand, a discrete Laplacian is not an operator of quantum geometry because it contains, beside geometric operators, maps acting on the function space of test fields. 
%Therefore, 
Note that only the coefficients of $\widehat\Delta_\cm$ 
are quantum operators in the usual sense, \ie maps from the Hilbert space $\mathcal{H_\cm}$ to itself.
$\widehat\Delta_\cm$ itself is an operator properly defined only on the coupled Hilbert space of geometry and test fields, which we do not introduce. 
We do not consider quantum states of test fields,
since the relevant object $\widehat {P(\tau)}$ to define the spectral dimension is a functional of pure geometry
%$\ds^{\psi}(\tau)$ of a quantum state of geometry $|\psi\rangle\in\mathcal H$
and, as such, it can eventually be defined as a quantum operator in the strict sense.
Let us expand this technical point for the interested reader. 
%A quantum operator in the usual sense is a map from the Hilbert space $\mathcal{H}$ to itself. 
As a vector space over complex numbers, any state in a Hilbert space can be expanded in the elements of a complete basis with complex numerical coefficients. 
Elements in the image of the quantum Laplacian $\widehat\Delta_\cm$ are sums over such a basis, but with coefficients that are discrete Laplacians instead of complex numbers, that is, maps from a functional space on a complex to itself. These elements are obviously not states in $\mathcal{H}$.
Still, we can use $\widehat\Delta_\cm$ to define $\widehat {P(\tau)}$ which is the quantum operator acting on pure-geometry states that we are interested in here.

%The Laplacian is itself a function on quantum-geometry states, inasmuch as it is a function of quantum-geometry operators (basically, the discrete metric). 
%\cjt{When acting on test fields, $\widehat\Delta$ returns a combination of test fields on the lattice rather than states of quantum geometry. 
%This is not a problem because we do not regard such test fields as defining quantum states of the theory. Therefore, the only operators we consider are those not depending on test fields, i.e., the Laplace operator itself and the return probability, the latter being defined as formal traces over fictitious test-field states. Of these operators we give the expression in terms of quantum-geometry observables and the action on quantum-geometry states. These objects are all that is needed for our purposes and are well defined.} 
}
 This can be done in different ways, depending on the geometric variables that are most convenient in the specific quantum geometric context that is chosen. 
It has been discussed in detail in \cite{\COTa}. 
In general, the resulting expression will be a complicated function of the quantum labels assigned to the complex, which is however both well-defined and explicitly computable \cite{\COTb}.

% clarification:
The operators $\widehat{P(\tau)}$ and $\widehat \Delta$  on the full Hilbert space $\mathcal H = \bigoplus_\cm \mathcal H_\cm$ are then defined in terms of the family of orthogonal projections $\pi_\cm:\mathcal H \ra \mathcal H_\cm$.
In this way, the Laplacian acting on generic quantum states of geometry is the formal sum
\[
\widehat \Delta := \sum_\cm \pi_\cm \widehat \Delta_\cm \pi_\cm \;.
\]
With the appropriate notion of a trace $\overline\Tr := \sum_\cm \Tr_\cm \pi_\cm$, based on the trace $\Tr_\cm$ over discrete field space on $\cm$ introduced above, the heat trace is then defined as
\[
\widehat{P(\tau)} := \overline\Tr\, \e^{\tau \widehat \Delta} \;.
\]
$\widehat{P(\tau)}$ is a map from $\mathcal H$ on itself, and thus a quantum operator in the strict  sense. 
Then, the spectral dimension $\ds^{\psi}(\tau)$ of a quantum state of geometry $|\psi\rangle\in\mathcal H$ is the scaling of the expectation value of  $\widehat {P(\tau)}$ \cite{\COTb}:
\[
\ds^{\psi}(\tau) := -2\frac{\partial}{\partial\ln\tau} \ln\langle \widehat{P(\tau)}\rangle _{\psi} %\nonumber\\
%&=& -2 \tau\partial_{\tau} \ln\langle \mbox{Tr}\widehat{K}(x,x';\tau)\rangle _{\psi}
.\label{eq:dSquantum}
\]
Note that it depends only on pure geometry, since the relevant operators are acting on pure-geometry states.

%clarification:
%The evaluation of the quantum operator $\widehat\Delta$ acting on quantum states of geometry $|\{j_c\},\cm\rangle$ with variables $\{j_c\}$ having support on a complex $\cm$ can be defined as the action of an operator version of these volume coefficients on $\cm$  \cite{\COTa}.

For the discrete quantum geometries $|\{j_c\},\cm\rangle$ it is reasonable to assume that they are eigenvectors of $\widehat \Delta_\cm$, based on the definition of these labels \eqref{spectra} and on our previous work \cite{\COTb}.
On the states \eqref{sups} that we are interested in here, %expanding 
the heat-trace expectation value is thus
%operator in the discrete quantum geometries $\rj$ gives
\begin{eqnarray}
\label{eq:dsexpand}
\langle \widehat{P(\tau)}\rangle _{\qs} & = & %\langle \qs| \widehat{ \Tr\, \e^{\tau\Delta}}|\qs\rangle = 
\sum_\cm \!
\left( \sumint_{j} \qsc^*_{j,\cm} \lj \right) \!\!\!\!
\left( \sum_{j'}  \qsc_{j',\cm} \, \Tr_\cm\, \e^{\tau\widehat{\Delta}_\cm} |j',\cm\rangle \right) \nonumber\\
& = & \sumint_{j,\cm} \left|\qsc_{j,\cm} \right|^{2} \Tr_\cm\, \e^{\tau \lj \widehat\Delta_\cm \rj}\,.
\end{eqnarray}
%where % in the last step 

%then leads to the expression 
%\begin{equation}
%\label{eq:dsbasis}
%\ds^{\qs}=-2\tau\frac{\underset{s}{\sum}\left|\qsc(s)\right|^{2}\mbox{Tr}\langle s|\widehat{\Delta}|s\rangle \rm e^{\tau\langle s|\widehat{\Delta}|s\rangle }}{\underset{s}{\sum}\left|\qsc(s)\right|^{2}\mbox{Tr}\,\rm e^{\tau\langle s|\widehat{\Delta}|s\rangle }}
%\end{equation}

% Assumption
Some simplifying assumptions are however needed in order to proceed with systematic computations on extended complexes. In the following, we assume that the expectation value of the %quantum-operator version of the 
Laplacian $\widehat\Delta_\cm$ %coefficients $\Delta_{ab}$ 
scales as
\[\label{assumption}
\lj \widehat\Delta_\cm \rj_{ab} \propto l^{-2}(j)\,(\Delta_\cm)_{ab}\,,
\]
where $\Delta_\cm$ is the combinatorial Laplacian \eqref{discrete-lap} on the complex $\cm$.
%(which is just the graph Laplacian of the 1-skeleton of the dual complex $\cm^\star$ \cite{\COTa}).
This assumption is sensible if the Laplacian can be expressed as a function of the volumes \eqref{spectra}. %, as it is indeed the case.
A similar Ansatz is, in fact, made in \cite{Modesto:2009bc}, although in that work this is not justified on the basis of a detailed analysis of the underlying graph and on the complete expression for the Laplacian, such as the one presented in \cite{\COTa}.

We now evaluate the spectral dimension on our superposition states. Under the assumption \eqref{assumption}, the expression for the expectation value of the return probability further simplifies to
\[
\langle \widehat{P(\tau)}\rangle _{\qs} \propto \sumint_{j,\cm} \left|\qsc_{j,\cm} \right|^{2} \Tr_\cm\, \e^{\tau l^{-2}(j) \Delta_\cm} .
\]

The above expression can be computed most efficiently considering the limit of infinite lattices, for which an analytic expression for the heat trace is available. In \cite{\COTb}, we showed that the heat trace on $\cminf = \Z^\sd$ is
\[
P^{\cminf}(\tau) = \left[\e^\tau I_0(\tau)\right]^\sd,
\]
where $I_0$ is the modified Bessel function of the first kind. In the same limit, one can give precise formulae for the contribution to the spectral dimension coming from individual lattices, so that we are in the ideal position to investigate the effect of superpositions of the same. The spectral dimension $\ds^{j, \cminf}$ on a single state $|j,\cminf\rangle$ equals $\sd$ for $\tau\gg l^2(j)$ and vanishes for $\tau \ll l^2(j)$:
\[\label{dslattice}
\ds^{j, \cminf}\simeq\left\{\begin{matrix} \sd\,,\qquad \tau\gg l^2(j)\\ 0\,,\qquad \tau\ll l^2(j)\end{matrix}\right.\,.
\]
Around the scale $\tau\approx l^2(j)$, there is a peak of approximate height $1.22 \sd$ \cite{\COTb}. We consider these features as discretization artefacts, and we conclude that no real dimensional flow is seen for individual states in the superposition \cite{\COTb}. 

Therefore, we are prompted to extend the search for quantum geometry states that would show true signs of dimensional flow to superposition states, to which we now move. Using the above solution, the spectral dimension of $\rsup$, Eq.\ \eqref{states} in the limit $N(\jmin)\ra\infty$, is given by the scaling of
\[\label{htsuperposition}
\langle \widehat P(\tau) \rangle_\ssup
\propto \sumint^\jmax_{j=\jmin} \left|\qsc_j \right|^2 \left\{\e^{l^{-2}(j) \tau} I_0[l^{-2}(j)\, \tau]\right\}^\sd .
\]
For asymptotic power-law spectra 
\[\label{beta}
l(j) \simeq j^\beta\,,
\]
where $\beta>0$ as usual in LQG [see Eq.\ \eqref{areaspectrum}], we have done numerical calculations for various classes of coefficient functions $\qsc_j$ and various values of spatial dimension $\sd$ and cutoffs $\jmax$. In all the examples presented here, we use $\jmin =1$; calculations with lower cutoffs of the same order (\eg  $\jmin =1/2$) give similar results. Notice also that the same finite minimal value for the geometric spectra could be obtained in correspondence with a quantum label $j=0$, for choices of quantization map that give a nonzero value for $C$ in Eq.\ \eqref{areaspectrum}. 

The first general class of coefficients to be considered is of power-law functions,
\[\label{gamma}
\qsc_j \propto j^\gamma \,.
\]
Defining the parameter
\[\label{alpha}
\alpha := -\frac{2\gamma+1}{\beta}\,,
\] 
the spectral dimension of the state under consideration has the following behaviour depending on the range of values of $\alpha$.
%In the critical dimension $d=\alpha$ the UV behaviour is not obvious from finite $\jmax$ calculations.

\begin{itemize}
\item For $0<\alpha<\sd$:
\begin{itemize}

\item[(a)] In the IR, \ie for large length scales $\tau\gg l^2(\jmax)$, $\ds(\tau)=\sd$ (Fig.\ \ref{SuperposD}). This is of course a consistency check for the validity of the formalism, since at large scales we recover the topological dimension of the space the quantum states are supposed to represent. It is however already a nontrivial test, as identifying quantum states with the right semiclassical continuum properties at large scales is no small task in background-independent quantum gravity. 

\begin{figure}
\includegraphics[scale=0.6]{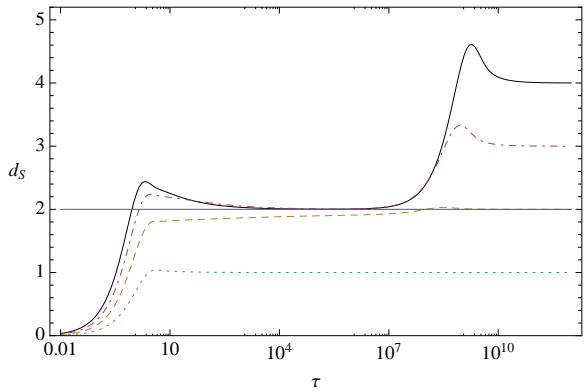}
\caption{Spectral dimension of a superposition with $\alpha = 2$ in $\sd=1,2,3,4$ (dotted, dashed, dot-dashed, solid curve) with cutoff $\jmax = 10^4\sd$.
}
\label{SuperposD}
\end{figure}

\item[(b)] Below the smallest lattice scale, \ie for $\tau\ll l^2(\jmin)$, $\ds(\tau) = 0$.  
This is the usual discreteness effect which we find also for individual lattice-based states \eqref{dslattice}, which remains at the Planck scale for discrete spectra induced by holonomies valued in compact groups \cite{DePietri:1996en,Ashtekar:1997bn,Achour:2014gr}. For noncompact groups, spectra are typically continuous and no volume discreteness effect at Planck scale occurs, as $\jmin \ra 0$ \cite{Freidel:2003kx}.

\item[(c)] Between these scales, there is a plateau with value $\ds(\tau) = \alpha$ (Fig.\ \ref{SuperposA}). This plateau indicates a regime in which the effective dimension is physically smaller than the topological one, and thus a proper dimensional flow. In the light of our previous results \cite{\COTb}, which, as discussed, were performed on the same type of quantum states and in the same formalism, but without considering large superpositions of lattice structures, we regard this as a truly quantum effect stemming from the superposition of states $\rj$ with geometric spectra on different scales and based on complexes of different size. It is interesting that, at such intermediate scales, the effective dimension is independent of the topological one (again, provided $\sd>\alpha$) and depends instead only on the specific choice of quantum states.

\begin{figure}
\includegraphics[scale=0.6]{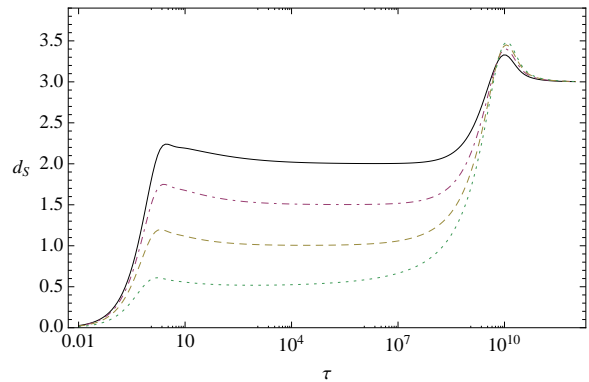}
\caption{Spectral dimension of superposition states with $\alpha = 1/2, 1, 3/2, 2$ (dotted, dashed, dot-dashed, solid curve) in $\sd=3$ with cutoff $\jmax = 10^5$.
}
\label{SuperposA}
\end{figure}

\item[(d)] In particular, for infinite superpositions ($\jmax\to\infty$) this plateau takes the value $\alpha$ and extends indefinitely (Fig.\ \ref{SuperposJ}). Formally, one can express this behaviour by
\[
\Delta\tau\big|_{\ds=\alpha} \underset{\jmax\to\infty}{\longrightarrow} \infty\,.
\]
Notice that this only means that the topological dimension $d$ is obtained further away at large $\tau$. Physically, one never takes the infinite limit in practice: for large spin labels, the plateau is long but has finite extension $\Delta\tau$.

\begin{figure}
\includegraphics[scale=0.6]{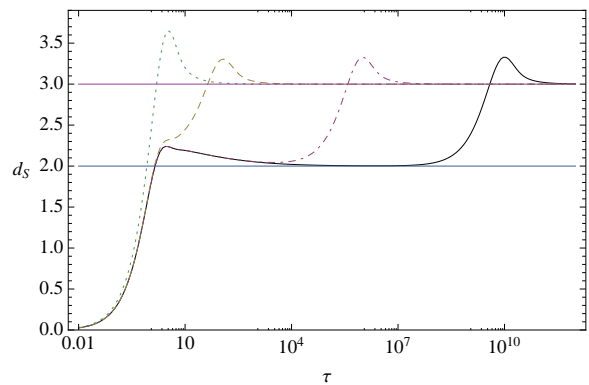}
\caption{Spectral dimension of superpositions with $\alpha = 2$ in $\sd=3$ for cutoffs $\jmax = 1,10,10^3,10^5$  (dotted, dashed, dot-dashed, solid curve).}
\label{SuperposJ}
\end{figure}

\item[(e)] Moreover, these results are independent of the spacing of the quantum labels $j$. 
Summing over $j\in\frac 1 q \N$ for some $q\in\Q$ slightly changes the results only at the scale $l(\jmin)$ (Fig.\ \ref{SuperposDiff}).
Therefore, neither the IR nor the UV regime depends on the spacing of the state label $j$. The numerical calculations show, in particular, that this should also be true in the limit $q\ra\infty$, \ie for positive real $j$.
% note that 
\end{itemize}

\begin{figure}
\includegraphics[scale=.6]{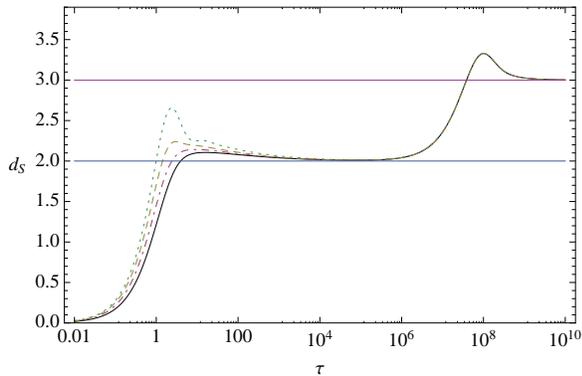}
\caption{Spectral dimension of a superposition with $\alpha = 2$ in $\sd=3$ summing over positive $j\in \frac 1 q \N$ up to $\jmax =  10^4$ for $q=1/2,1,2,10$  (dotted, dashed, dot-dashed, solid curve).
}
\label{SuperposDiff}
\end{figure}

\item For $\alpha<0$, no superposition effect occurs and the profile of the spectral dimension equals approximately the one of the single state $|\jmax,\cminf\rangle$, Eq.~(\ref{dslattice}):
\[
\ds^{\ssup}(\tau) \approx \ds^{\jmax,\cminf}(\tau).
\]
This is a numerical result, for which we lack, at present, a complete analytical or physical understanding. Nevertheless, we can offer an intuitive explanation. We saw that, in the range $0<\alpha<d$, $\alpha$ is the spectral dimension of the state at sufficiently small scales. On a continuous medium, the case $\alpha<0$ would correspond to an unphysical one with negative dimension. This situation is meaningless both in the conventional diffusion interpretation of the spectral dimension (where the probe would do ``less than not propagating'') and in the resolution interpretation of \cite{CM1,CMN}. In the latter, the return probability $P(\tau)\sim (\sqrt{\tau})^{-\ds}\sim \ell^{-\ds}\sim ({\rm res})^{\ds}$ is the probability to find the probe anywhere when the geometry is probed at scales $\ell$, i.e., with resolution $1/\ell$. For positive $\ds$, this probability decreases with the resolution: if $1/\ell$ is too small, there is a chance that we do not see the probe at all. On the other hand, a negative $\ds$ implies that the coarser the probe, the greater the chance to find it somewhere. In our case, this pathological behaviour is screened by discreteness effects and $\ds$ is saturated by the lattice with labels $\jmax$. The resolution interpretation coupled with the LQG interpretation of the spin labels helps in explaining Eq.\ (\ref{dslattice}): coarser resolutions can effectively probe only large volumes and the largest volume available for the states (\ref{htsuperposition}) is at the upper cutoff $\jmax$. Under such conditions, it is natural to expect that the lattice structure completely dominates the profile of the spectral dimension. 

\item For $\alpha>\sd$, no superposition effect occurs and the profile of the spectral dimension equals approximately the one of the single state $|\jmin,\cminf\rangle$,
\[\label{dsmin}
\ds^{\ssup}(\tau) \approx \ds^{\jmin,\cminf}(\tau).
\]
In the continuum limit, $\alpha>d$ would imply a spectral dimension larger than the ambient space. Similarly to the previous case, one has both the diffusion and the resolution interpretation at hand. In the conventional diffusion interpretation of the spectral dimension, the case $\ds>d$ may be regarded as physical: the probe effectively sees more than $d$ dimensions and tends to superdiffuse. In the resolution interpretation, the probability of finding the probe somewhere grows more steeply than for the normal case (Brownian motion) and probes with large resolution (small scales $\ell$) become even more effective. However, in the present quantum framework there is a limit to which one can probe the microscopic structure of geometry: volume spectra are discrete with minimum eigenvalue determined by $\jmin$. Again, the variation of the spectral dimension is dominated by lattice effects, this time governed by the lower cutoff in the spin labels.
\end{itemize}
% Discussion on technical aspects of the results
A partial understanding of the results with $0<\alpha<d$, in particular concerning the dependence of the UV value of $\ds$ on the powers $\beta$ and $\gamma$ in \eqref{alpha}, is provided by the following rewriting of the heat trace \eqref{htsuperposition}.
%characterized by a particular set of coefficients {\bf [a justification for this choice of coefficients is missing and it is one of the main points of the paper]}
%\[\label{coefficients}
%\qsc_\alpha(j) := \sqrt{\frac \d {\d j} \left[\frac {l(j)}{l_0}\right]^\alpha}
%=  \sqrt{\alpha \frac{\d\ln l(j)}{\d j}  \left[\frac {l(j)}{l_0}\right]^\alpha}.
%\]
%Thereby, a state $\rsup$ is completely determined by a parameter $\alpha\in\R$ and the volume spectrum $l(j)$, \eqref{spectra}.
A redefinition of variables $k(j) := l^{-\alpha}(j)$ demands a change of summation-integration measure by %the squared modulus of the expansion coefficients of $\rsup$ up to a sign:
\[
\frac{\d k}{\d j}  = \frac \d {\d j} l^{-\alpha}(j)
=  -\alpha \frac{\d\ln l(j)}{\d j}  l^{-\alpha-1}(j)\,.
%\pm \left|\qsc_\alpha(j)\right|^2\,,
\]
In particular, for the power-law spectra \eqref{beta} and the definition of $\alpha$ \eqref{alpha}
\[
\frac{\d k}{\d j} =  -\alpha \beta \, j^{-\alpha \beta -1} \os{\text{\scriptsize \eqref{alpha}}}{=} (2\gamma+1)\, j^{2\gamma}
\]
which is proportional to $|\qsc_j|^2$ for the power-law coefficients \eqref{gamma}.
Thus, the heat trace on these superpositions is a uniformly weighted sum in the $k$-variable [over the range corresponding to \eqref{htsuperposition}]:
\[\label{ksum}
\langle \widehat P(\tau) \rangle_\ssup  \propto  \sumint_k \left[\e^{-k^{2/\alpha}\tau} I_{0}({ k^{2/\alpha}}\tau)\right]^{d}.
%P_{\Z^{d}}\left(\left(\frac{r}{r_{diff}}\right)^{2\alpha}\tau\right)\\
%& = & \sum_{r=r_{min}}^{r_{max}}e^{-d\left(\frac{r}{r_{diff}}\right)^{2\alpha}\tau}\left\{ I_{0}\left(\left(\frac{r}{r_{diff}}\right)^{2\alpha}\tau\right)\right\} ^{d}
\]
Therefore, genuine dimensional flow comes from a subtle balancing of $\sd$ and $\alpha$ in this expression, while a negative $\alpha$ yields just a dominant $\kmax = k(\jmax)$ contribution in the sum. Indeed, we have also calculated the spectral dimension directly from \eqref{ksum} for various values of $\sd$, $\alpha$ and summing ranges of integer $k$'s, obtaining qualitatively similar results as discussed above for \eqref{htsuperposition}. As a consequence, dimensional flow has some dependence on the form of the spectrum \eqref{beta} but only in combination with appropriate superposition coefficients.

\ 

Still maintaining the power-law spectrum \eqref{beta} (which is the most reasonable assumption, consistent with known results in LQG and related approaches), we have calculated the spectral dimension for various other classes of coefficient functions. In most cases, there are no surprising results.

(a) For example, exponential coefficients $\qsc_j\propto \e^{aj}$ let either the maximal state $\jmax$ dominate when $a>0$, or the minimal one $\jmin$ when $a<0$. 
(b) Gaussian coefficients, on the other hand, result in a dominance of the $j_0$ at which they are peaked. (c) Trigonometric functions add some oscillations to $\ds^{\jmax,\cminf}$ in the intermediate regime, depending on the relation of the periods to the spacing of $j$ in the sum. In all these cases, therefore, the overall behaviour of the spectral dimension is the same as that found for coefficients given by simple powers. 

\begin{figure}
\includegraphics[scale=0.6]{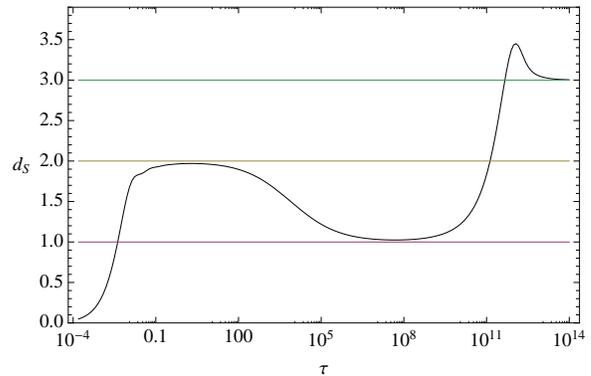}
\caption{Spectral dimension of superpositions with coefficients $|c_j|^2 = j^{-4} + 200 j^{-7}$ summing from $\jmin =1/2$ to $\jmax = 200$ for $\sd=3$ and $\beta = 3$ (to be able to numerically cover enough scales with a feasible number of states in the sum). According to \eqref{alpha}, two different UV regimes with dimension $\ds\approx 2$ and then $\ds\approx 1$ can be observed.
}
\label{regimes}
\end{figure}

More interesting is the case of coefficients which are linear combinations of power functions in $j$. Then one finds, for their asymptotic behaviour $\qsc_j\sim j^\gamma$, the same effect as for power-law coefficients. In particular, if there are several regimes with different approximate scaling $\gamma_1, \gamma_2, \dots$, one obtains plateaux in the spectral dimension plot of different values $\alpha_1, \alpha_2, \dots$ accordingly. An example is shown in Fig.\ \ref{regimes}. This effect coincides, both in its qualitative shape and origin, to the one obtained in the multiscale generalization of the diffusion equation with different powers of the Laplacian \cite{frc4}. In general, all coefficient functions with an approximate power-law behaviour in some regime give rise to dimensional flow at those scales. Details such as the value of $\jmin$ and the spacing in $j$ are not relevant for the value of the spectral dimension in these intermediate regimes, in agreement with the discussion in \cite{frc4} on the role of regularization parameters in the profile of $\ds$. The details of regularization schemes are nonphysical and affect only transient regimes in $\ds(\tau)$, not the value of the plateaux.

%One can then conjecture that, generically, all coefficient functions with an approximate power function behaviour in some regime have a dimensional flow at the these scales. %according to the value of $\alpha$ \eqref{alpha}. 
%The case for the validity of such conjecture is strengthened further by the fact, already explained, that the details (value of $\jmin$, spacing in $j$) are not relevant for the value of the spectral dimension in the intermediate regime, where, as we have discussed, interesting new behaviour is expected.

% Calculation \dw ===============================================================

\subsection{Walk dimension of superpositions}

The spectral dimension is only one of the possible dimensional observables. Our strategy is well suited to analyze other observables as well, and it is interesting to do so because there exist several relations among them, in classical and continuum spaces. Only a detailed analysis of their combined behaviour can give solid indications on the nature of the quantum geometries corresponding to quantum gravity states.

A closely related observable is the walk dimension $\dw$. It is defined via the scaling of the mean square displacement
\begin{equation}
\langle X^{2}\rangle _{y}(\tau)=\int \d x \left|x-y\right|^{2}K(x,y;\tau)\propto\tau^{2/\dw},
\end{equation}
that is 
\begin{equation}
\dw(\tau):=2\left(\frac{\partial\ln\langle X^{2}\rangle _{y}}{\partial\ln\tau}\right)^{-1}.
\end{equation}
%If the space is not translation invariant, one can include an extra space average by integrating over spatial points $y$. {\bf [I'm not sure about this, it could give a trivial result (check the same trick for the Hausdorff dimension)]}

In the case of the $\sd$-dimensional hypercubic lattice $\cminf$, we can choose the origin $y=0$, so that
\begin{eqnarray}
\langle X^{2}\rangle _{0}^{\cminf}(\tau) & = & \sum_{\vec{n}\in\Z^{d}}\left|\vec{n}^{2}\right|K(\vec{n},0;\tau) \\
& = & \sum_{\vec{n}\in\Z^{d}}\left(\sum_{j=1}^{d}n_{j}^{2}\right)\e^{-\tau}\prod_{k=1}^{d}I_{n_{k}}(\tau)\,.
\end{eqnarray}
This can be evaluated using standard relations of the Bessel functions $I_n$,
\begin{eqnarray}
\langle X^{2}\rangle _{0}^{\cminf}(\tau) & \propto & \e^{-d\tau}\sum_{j=1}^{d}\sum_{\vec{n}\in\Z^{d}}n_{j}^{2}\prod_{k=1}^{d}I_{n_{k}}(\tau)\\
 & = & \e^{-d\tau}\sum_{j=1}^{d}\left[\sum_{n_{j}\in\Z^{d}}n_{j}^{2}I_{n_{j}}(\tau)\right]\nonumber\\
&&\times\prod_{k\ne j}^{d}\left[\sum_{n_{k}\in\Z}I_{n_{k}}(\tau)\right] \nonumber\\
 & = & \e^{-d\tau}d\left[\frac{\tau}{2}\sum_{n\in\Z}I_{n-1}(\tau)+I_{n+1}(\tau)\right]\left(\e^{\tau}\right)^{d-1} \nonumber\\
 & = & d\,\tau\,.
\end{eqnarray}
Thus, the walk dimension on the lattice is
\begin{eqnarray}
\dw^{\cminf}(\tau)=2\,,
\end{eqnarray}
as in the continuum.

Quantum superpositions $\rsup$ are characterized by the Laplacian \eqref{assumption}, so that along the same lines as \eqref{htsuperposition} one has
\begin{eqnarray}
\left\langle \langle X^{2}\rangle_0 (\tau) \right\rangle_\ssup 
&=& \sumint^\jmax_{j=\jmin} \left|\qsc_j \right|^2 {\langle X^{2}\rangle}^\cminf_{0}[l^{-2}(j)\tau]\nonumber \\
&=& d \sumint^\jmax_{j=\jmin} \left|\qsc_j \right|^2 l^{-2}(j)\tau\label{dwtem} \\
&\propto& \tau\,.
\end{eqnarray}
Therefore, also for quantum superpositions the scaling of the mean square displacement
%\begin{equation}
%\frac{\partial}{\partial\ln\tau} \ln \lsup \langle X^{2}\rangle _{0} \rsup
%= \tau\frac{\sumint_k l_0^{-2} k^{-2/\alpha}}{\sumint_k l_0^{-2} k^{-2/\alpha}\tau}=1\,,
%\end{equation}
yields the standard result
\[\label{dwfinal}
\dw^\ssup=2\,,
\]
independent of the form of the coefficients $\qsc_j$. Notice that the dependence on the topological dimension in Eq.\ (\ref{dwtem}) is only through a proportionality coefficient. Therefore, Eq.\ (\ref{dwfinal}) is valid both for space and spacetime.

% Calculation \dh ===============================================================

\subsection{Hausdorff dimension of superpositions}

The Hausdorff dimension of a quantum state is defined in terms of the scaling of the expectation value of the volume $V(R)$ of a ball with radius $R$:
\[
\dh^{\qs}(R):=\frac{\partial\ln \langle V(R) \rangle_{\qs}}{\partial\ln R}\,,
\]
which can be further expanded like the spectral dimension \eqref{eq:dsexpand}.
Using the graph distance and measuring $R$ in units of the lattice spacing, the volume on the lattice $\cminf$ is
\[\label{Hausdorffproperties}
V^{\cminf}(R)= 2^{d}\left(\begin{array}{c}
R+\sd-1\\
\sd
\end{array}\right)
= \frac{2^\sd}{\sd!}\prod_{n=0}^{\sd-1} (R+n)\,,
\]
yielding the Hausdorff dimension
\[\label{dh1}
\dh^{\cminf}(R)=R\sum_{n=0}^{d-1}\frac{1}{R+n}=R\left[\digamma(R+\sd)-\digamma(R)\right]\,,
\]
 where $\digamma$ is the digamma function. At large scales, $\dh$ approaches the topological dimension $d$, while at small scales it tends to 1:
\[\label{dh2}
\dh^{\cminf} \simeq\left\{\begin{matrix}\sd\,,\qquad R\gg1\\ 1\,,\qquad R\ll1\end{matrix}\right..
\]
%$ \dh(R\gg1) %\us{R\ra \infty}{\lora} \sum_{k=0}^{d-1} 1
% = \sd$, while at small scales $\dh(R\ll1) = %1 + \sum_{k=1}^{d-1}\frac{R}{R+k} \us{R\ra 0}{\lora}  
%1$.

On discrete quantum geometries $\rjc$, we define the quantum analogue of $V(R)$ as follows. Let $v_0\in\cm$ be a given vertex in the complex and consider the subcomplex $\cm_{v_0}\In\cm$ of all vertices $v$ which have an expectation value of their distance to $v_0$ no larger than the radius, 
\[\label{radius}
\ljc \widehat{L}_{vv_0}\rjc \le R\,,
\]
where the expectation value of
$
\widehat{L}_{vv_0} %:= \text{Min}\{ 
$
is the minimum of lengths derived from the sum of edge lengths of possible (combinatorial) paths between $v$ and $v_0$.

The expectation value of the volume of this subcomplex $\rjc$ %on a state $|\psi\rangle$ 
is $\sum_{v\in\cm_{v_0}}\langle V_v \rangle_{\{j_c\},\cm%\qs,\cm_{v_0}
}$. To obtain the desired observable, one must average over all possible centers $v_0$:
\[
\ljc V(R) \rjc  = \sum_{v_0\in\cm} \sum_{v\in\cm_{v_0}} \ljc V_v \rjc\,. %\rangle_{\qs,\cm_{v_0}}.
% = \sum_{v_0\in\cm} \sum_{\us{\langle l_{vv_0} \rangle \le R}{v\in\cm}} \langle V_v \rangle_{\qs,\cm}.
\]
On the uniform hypercubic lattice states $\rj$, however, the sum over center vertices $v_0$ is not necessary due to translation invariance and because of the local volumes being all equal, $\lj V_v \rj \propto l^{\sd}(j)$ for all $v\in\cminf$.

Similarly, on $\rj$ the condition \eqref{radius} simplifies to 
\[
\lj \widehat{L}_{vv_0} \rj \propto l(j) N_{vv_0}
\]
where now $ N_{vv_0}$ is the minimal number of edges in a path from $v$ to $v_0$.

\begin{figure}
\includegraphics[scale=.6]{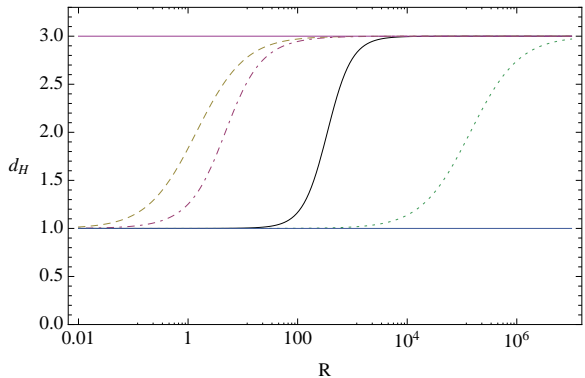}
\caption{Hausdorff dimension $\dh$ of a superposition with $\alpha = 1,2$ (solid and dash-dotted curve) in $\sd=3$ summing up to $\jmax =  10^5$, compared to $\dh$ on single states $|1,\cminf\rangle$ (dashed curve) and $|\jmax,\cminf\rangle$ (dotted curve).
}
\label{Hausdorff}
\end{figure}

Therefore, the evaluation of $V(R)$  on $\rj$ can be expressed in terms of $V^\cminf (R)$ as 
\begin{eqnarray}
\lj V(R) \rj &\propto& l^\sd(j) V^\cminf [R/l(j)]\nonumber \\
&\propto& l^\sd(j) \prod_{n=0}^{\sd-1} [R /{l(j)}+n]\,.
\end{eqnarray}
As for the spectral dimension \eqref{htsuperposition}, this gives a nontrivial expectation value for generic superposition states:
\[\label{vrsuperposition}
\langle V(R) \rangle_\ssup \propto \sumint_{j=\jmin}^\jmax |\qsc_j|^2 l^\sd(j)  \prod_{n=0}^{d-1}[R /{l(j)}+n]\,.
\]
Nevertheless, numerical calculations on the same classes of states as investigated above for the spectral dimension show qualitatively similar results to the Hausdorff dimension $\dh^{j,\cminf}$ on single states $\rj$ (Fig.\ \ref{Hausdorff}). 
That is, in all instances there are plateaux as in the pure lattice case \eqref{Hausdorffproperties}.
Only the scale and steepness of the flow between these plateaux is modified. 
For example, for power-law coefficients \eqref{gamma} the fall-off is much steeper and occurs, as $\alpha$ increases, closer to the scale as in the case of the single state $|\jmin,\cminf\rangle$.

%\cjt{this discussion might be extended, but not necessarily so if there is really not more to say also for the other classes of coefficients...}

%Thus, there are no deviations from the Hausdorff dimension (\ref{dh1})-(\ref{dh2}) of the classical lattice.

% Discussion ==================================================================

\section{Discussion}

%\subsubsection*{Discussion}

Our calculations have shown that a flow in the spectral dimension occurs in quantum gravity, at least for a specific class of superpositions of regular (both from the combinatorial perspective and for what concerns the assignment of additional quantum labels) quantum states of geometry. These quantum states, although restricted by the regularity assumption, are exactly of the type appearing in the related quantum-gravity formalisms of loop quantum gravity, spin-foam models and group field theory, but can also simply be seen as quantum states of lattice quantum gravity, in the spirit of quantum Regge calculus. 

On the other hand, we see no dimensional flow due to quantum effects in the Hausdorff and walk dimension. This conclusion is based on the interpretation, which we maintained throughout the paper, that the flow of a geometric indicator is an artefact of discretization effects whenever it approximately coincides with the flow for lattices. We will come back to this point.

Let us comment a bit further about our results from the point of view of loop quantum gravity.

Under the assumptions made for the action of the quantum Laplacian on the states (a very simple scaling behaviour), an important example of states of the type we have studied are kinematical states in LQG where length ($\sd+1=3$) or area and volume operators ($\sd+1=4$) are diagonalized by spin-network states. 
%\cjt{should we make more explicit, that the Laplacian is assumed to be diagonal as well?}
In this sense, we have identified a class of LQG states with a dimensional flow. 
More precisely, for any $0<\alpha<\sd$ there is a class of states in the kinematical Hilbert space with a dimensional flow from the spatial topological dimension $\sd$ in the IR to a smaller value $\alpha$ in the UV. The UV value depends on the exact superposition considered but not on the topological dimension.%, in a regime around the scale set by the cutoff $\jmax$ in the superposition.

This result is in contrast with earlier arguments in LQG \cite{Modesto:2009bc}. There, the author argues for evidence of dimensional flow for individual spin-network states (thus, for a given graph or complex), and the same result is claimed in \cite{Caravelli:2009td,Magliaro:2009wa} for simple spin-network states with additional weights given by a 1-vertex spin foam (thus, not yet in a truly dynamical context). The starting point in \cite{Modesto:2009bc} is an assumption about the scaling of the expectation value of the Laplacian, very similar to \eqref{assumption}. The essential part of the argument is then the further assumption that the momenta $p$ of the scalar field defining the spectral dimension are directly related to a length scale set by the quantum numbers as $p\propto1/\sqrt j$. The scaling of the Laplacian in $j$ is then translated into a modified dispersion relation in $p$ and the result depends on the precise form of the spectrum \eqref{areaspectrum} with $C=0$.

In our case, no further assumption beyond \eqref{assumption} is made. Calculations are based on the momenta of the scalar field on the lattice-based geometry, that is, the spectrum of the Laplacian, but the spectral dimension is computed directly as a quantum geometric observable evaluated on quantum states. As recalled already above, in a previous work using this more direct approach \cite{\COTb} we have found no effects on the spectral dimension for individual quantum-geometry states of LQG based on given graphs or complexes. On the other hand, the genuine dimensional flow that we have encountered here for the states $\rsup$ is crucially related to the superposition of spin-network states also with respect to the underlying combinatorial structures, and it is not solely the result of the discreteness of geometric spectra. In this deeper sense, dimensional flow can indeed be seen as an effect of quantum discreteness of geometry.

%We can further stress this point by considering a simplified version of the states (\ref{state}). Let the spectra have a power-law dependence on the quantum labels, $l(j)= l_0 j^\beta$. The defining coefficients \eqref{coefficients} then becomes
%\[
%\qsc_\alpha(j) = \left[\frac \d {\d j} j^{\alpha\beta} \right]^{\frac 1 2}
%=  \sqrt{\alpha\beta} j^{\frac{\alpha\beta-1}{2}}.
%\]
%With this Ansatz, states $|1/\beta\rangle$ are uniformly weighted superpositions, \ie, $\qsc_{1/\beta}(j)= \qsc_{1/\beta}= {\rm const}$. If $\beta$ were negative, these states would have a dimensional flow to $|1/\beta|$. However, this is not the case in any known theory of quantum geometries. In the particular case of LQG, $\beta$ is positive for lengths and areas in $\sd=3$ and $\sd=4$, as mentioned above. 

%This dimensional flow is not a discreteness effect in the sense of being related to discrete quantum spectra (property \eqref{assumption}). Still, single states are discrete in that they are based on combinatorial manifolds, in the case discussed here hypercubic lattices. 
\

We are also in a position to characterize the change of dimensionality more precisely than a generic ``flow'' of geometry. Quite often in the literature of quantum gravity, dimensional flow has been advertised as spacetime being ``fractal.'' However, strictly speaking not all sets with varying dimensionality are fractals. Although no unique operational and rigorous definition of fractal exists, one property all fractals generally possess is a special relation among the spectral dimension $\ds$, the Hausdorff dimension $\dh$ and the walk dimension $\dw$:
\begin{equation}
\dh=\frac{\dw}{2}\ds.\label{eq:dimensions}
\end{equation}
%If the quantum geometries we considered satisfied this relation non-trivially (i.e., in the presence of genuine dimensional flow, where at least one of the dimensions acquire scale dependence), then we could tentatively conclude that they are fractal in the strict sense of (\ref{eq:dimensions}).

On the hypercubic lattice superpositions that we have considered, $\dw=2$ and the above relation simplifies to $\dh=\ds$. This is trivially obtained in the IR regime, where both dimensions take the value of the topological dimension. In the UV regime above the lattice scale (recall that below such scale any scaling effect is arguably unphysical), the Hausdorff dimension takes the classical value $\dh^{\cminf}\simeq 1$. Thus, \eqref{eq:dimensions} is only obeyed in the case of a scaling $\alpha = 1$ such that also the spectral dimension takes this value. Only then can one call the quantum superposition $\rsup$ an effective one-dimensional fractal. This is indeed a perfectly allowed choice of quantum states and we can conclude that we have identified a particular class of quantum geometries that corresponds, by all appearances, to a fractal quantum space.

However, we should mention a caveat here. For these geometries to be safely regarded as fractals, the origin of the dimensional flow should be the same in the left- and right-hand side of Eq.\ \eqref{eq:dimensions}, which may not be the case for us: the left-hand side flows due to discreteness effects, while the right-hand side flows due to physical quantum effects. This situation might suggest either that we should not place particular significance in the fulfillment of Eq.\ \eqref{eq:dimensions} or that
our discrimination between discreteness artefacts and physical effects is somewhat too strong and should be revised. We no dot attempt to solve this mild conceptual issue here, which is harmless for our main results. Still, it will deserve further attention.

Interestingly, the geometry with $\alpha=1$ is also the only one where the spectral dimension of spacetime reaches the value $\Ds=\ds+1\simeq 2$ so often commented upon in the literature of quantum gravity. Its appearance across independent approaches such as causal dynamical triangulations, asymptotic safety, Ho\v{r}ava--Lifshitz gravity and others \cite{tHo93,Car09,fra1} triggered the suspicion that this ``magic number'' was a universal characteristic of frameworks with good ultraviolet properties or, in other words, that a two-dimensional limit of the spectral dimension was tightly related to the renormalizability or finiteness of quantum gravity. By now, it has become clear that this is not the case in general, as there exist counterexamples of nonlocal field theories with good renormalization properties \cite{Biswas:2006jn} with $\Ds\neq 2$ in the UV \cite{Mod1}, as well as of local theories whose renormalization properties are not at all improved by dimensional flow \cite{frc9}. Here we provide another instance pointing towards the same conclusion: the value of $\ds$ is governed by a choice of states which, by itself, is not (sufficiently) connected with the dynamical UV properties of the underlying full theory.

% Conclusions ==================================================================

\section{Conclusions}

We have investigated the effective structure of quantum superpositions of regular (hypercubic and homogeneous in label assignment) states of quantum geometry. 

It is possible to identify states with a flow of the spectral dimension to a dimension $\alpha$ in the UV, provided the superposition includes fine enough combinatorial structures and a large enough number of (kinematical) degrees of freedom of quantum geometry, and a particular set of expansion coefficients \eqref{gamma} related to $\alpha$ \eqref{alpha}.

For the Hausdorff and walk dimension, no physical quantum effects are observed, although discreteness effects do alter the value of the Hausdorff dimension across scales.
A fractal structure in the strict sense, \ie where \eqref{eq:dimensions} relating the three dimensions is fulfilled also in the UV, is realized in the case $\alpha=1$.

In particular, these results provide evidence for a dimensional flow in a certain class of kinematical LQG states, also available in the spin-foam and group field theory context.

The results at hand can be further generalized in various directions as well as refined within individual theories of quantum gravity. In parallel, it becomes feasible to explore the phenomenological consequences of the discovered dimensional flow and (when applicable) fractal nature of quantum space as a direct effect of the full theory. This possibility is especially interesting in a quantum cosmological context, where a change of dimensionality can bear its imprint in the early stage of cosmic evolution \cite{AAGM1,AAGM2,frc11}. 

% ===========================================================================

\providecommand{\href}[2]{#2}
%\begingroup
\raggedright
%\endgroup

\end{document}